\documentclass[a4paper, conference]{IEEEtran}
\makeatletter
\ifCLASSOPTIONtwocolumn
\fi
\makeatother

\IEEEoverridecommandlockouts

\usepackage{cite}
\usepackage{amsmath,amssymb,amsfonts}
\usepackage{graphicx}
\usepackage{textcomp}
\usepackage{xcolor}
\usepackage{booktabs}
\usepackage{array}
\usepackage{algorithm}
\usepackage{algpseudocode}
\usepackage{color}
\usepackage{subcaption}
\usepackage[bookmarks=false]{hyperref} 

\def\BibTeX{{\rm B\kern-.05em{\sc i\kern-.025em b}\kern-.08em
    T\kern-.1667em\lower.7ex\hbox{E}\kern-.125emX}}
\begin{document}

\title{Anti-Jamming based on Beam-Steering Antennas and Intelligent UAV Swarm Behavior\\
\thanks{This work is funded by national funds through FCT — Fundação para a Ciência e a Tecnologia, I.P., under projects/supports UID/6486/2025 (https://doi.org/10.54499/UID/06486/2025) and UID/PRR/6486/2025 (https://doi.org/10.54499/UID/PRR/06486/2025).}
\thanks{© 2026 IEEE. Personal use of this material is permitted. Permission from IEEE must be obtained for all other uses, in any current or future media, including reprinting/republishing this material for advertising or promotional purposes, creating new collective works, for resale or redistribution to servers or lists, or reuse of any copyrighted component of this work in other works.}
}

\author{\IEEEauthorblockN{1\textsuperscript{st} Tiago Silva}
\IEEEauthorblockA{\textit{Military Academy} \\
\textit{Portuguese Army}\\
Lisbon, Portugal \\
silva.tas@exercito.pt}
\and
\IEEEauthorblockN{2\textsuperscript{nd} António Grilo}
\IEEEauthorblockA{\textit{INESC INOV and INESC-ID} \\
\textit{Instituto Superior Técnico, University of Lisbon}\\
Lisbon, Portugal \\
antonio.grilo@inov.pt}
}

\maketitle

\begin{abstract}
In recent years, Unmanned Aerial Vehicles (UAVs) have brought a new true revolution to military tactics. While UAVs already constitute an advantage when operating alone, multi-UAV swarms expand the available possibilities, allowing the UAVs to collaborate and support each other as a team to carry out a given task. This entails the capability to exchange information related with situation awareness and action coordination by means of a suitable wireless communication technology. In such scenario, the adversary is expected to disrupt communications by jamming the communication channel. The latter becomes the Achilles heel of the swarm. While anti-jamming techniques constitute a well covered topic in the literature, the use of intelligent swarm behaviors to leverage those techniques is still an open research issue. This paper explores the use of beam-steering antennas and UAV swarm formation adaptation in order to mitigate the effect of jamming in the main coordination channel, under the assumption that a more robust and low data rate channel is used for formation management signaling.  A Genetic Algorithm (GA) approach is used to jointly optimize beam directions, swarm formation and traffic routing. Simulation results show the effectiveness of proposed approach. However, the significant computational cost paves the way for future research.

\end{abstract}


\section{Introduction}
Unmanned Aerial Vehicles (UAVs) are increasingly employed in civilian and military operations due to their versatility and autonomous capabilities. In military contexts, UAVs are particularly valuable for reconnaissance, combat, and logistical support, providing an edge in information gathering and rapid response \cite{kim2017real, siddiqui2017development, brust2015networked}. However, UAV communications are highly vulnerable to jamming, which can significantly reduce their effectiveness on the battlefield. Like frequency hopping, existing countermeasures often must catch up in dynamic, high-speed data environments, particularly within UAV swarms where coordinated responses are essential.

This paper addresses the challenge of maintaining communication within a swarm of UAVs under jamming conditions. We propose a novel anti-jamming framework that combines advanced UAV antenna models with intelligent formation management. The primary research question of this paper is the following: 

\textit{Can a UAV swarm, acting in coordinated formation, leverage the effectiveness of anti-jamming techniques by jointly optimizing communication parameters such as formation, smart antenna configuration, and network traffic routing, without compromising the mission?}

The main contributions of this work are the following:
\begin{enumerate}
	\item The definition of a novel optimization problem, which seeks to jointly optimizing the swarm formation, smart antenna configuration, and network traffic routing, to achieve more effective anti-jamming performance. 
	\item A Genetic Algorithm (GA) approach to show how optimizing UAV positions and antenna beam directions can maximize communication capacity in the presence of interference and/or malicious jamming.
\end{enumerate}

This framework advances UAV swarm research, offering practical tools to improve resilience of multi-UAV teams and swarms against jamming in contested environments.  

\section{Related Work}
UAVs have transformed military and civilian applications, ranging from reconnaissance and surveillance to active combat roles. In military scenarios, UAVs offer the advantage of rapid deployment and high maneuverability, allowing them to perform missions such as reconnaissance, communication relay, combat support, etc. However, UAV communication systems are particularly susceptible to jamming attacks, where adversaries interfere with signal transmission to disrupt command and coordination efforts \cite{li2021networked, udeanu2016unmanned}.

Jamming countermeasures \cite{mpitziopoulos2009survey} often involve techniques like frequency hopping and physical path adjustments. These approaches have limitations, particularly in scenarios that require high data transmission rates and coordination within a UAV swarm. Swarm control allows multiple UAVs to coordinate formation, movements and communication strategies to more effectively mitigate the jamming impact. This is critical when the location and behaviour of the jammer are unpredictable, making static countermeasures less effective.

Previous research exists on swarm formation and movement adaptation to mitigate jamming and interference.

In \cite{liang2018}, mobile devices exploit two-dimensional mobility to deal with interference. The effects of introducing specific parameters into a Q-learning algorithm were investigated. The parameters selected for analysis were the frequency and mobility of individual nodes. Applying
these parameters to the Q-Learning algorithm resulted in a significant acceleration of the iterative process and greater
resistance to interference.


In \cite{peng2019anti}, the authors propose a Q-Learning algorithm for system optimization. The scenario in question involves communication between a swarm of UAVs and a ground station through a relay UAV. This relay optimizes the quality of communication in the receiving area by exploiting different parameters, such as the frequency, movement and spatial domain of the antenna. A cost metric is introduced to balance the movement and communication performance of the relay. Simulation results demonstrate the effectiveness of the proposed algorithm in improving UAV swarm communication
under interference. 

In \cite{wei2022}, the authors leverage Artificial Potential Fields (APFs) to improve communication between UAVs and ground stations in environments affected by suppression jamming. Using the  method, the authors leverage cooperative behavior among multiple UAVs. The system constructs communication potential fields based on the Received Signal Strength Indicator (RSSI) and Signal-to-Interference-plus-Noise-Ratio (SINR) values — both along the UAV cluster trajectory and at its location, assuming stable intra-cluster communication. These fields help estimate the direction of an unknown jamming source. A repulsive potential is then assigned to the jammer, while an attractive one guides toward the target. The APF algorithm generates a flight path that optimizes communication quality. Simulations validate the approach by showing reduced communication loss under interference.

In \cite{jeong2024anti}, the authors use a Graph Convolutional Network (GCN) to predict the location and intensity of interference areas based on the information collected by each UAV. A multi-agent control algorithm is then used to disperse the swarm, avoid the
interference areas and allow the UAVs to regroup when they reach their destination. This research stands out for scenarios where the position of the jammer is unknown, focusing on predicting
the areas of interference based on the collective intelligence of the UAVs, making it more robust and applicable to
real scenarios. The article concludes by demonstrating the effectiveness of the approach through simulations, where the
UAVs could avoid interference and achieve their objectives efficiently.

In \cite{xue2024}, the authors propose an anti-jamming attack mixed strategy for formation tracking control of the multi-UAV system. There are three types of UAVs: leaders, followers, and jammers. A three-layer Stackelberg game is constructed. Leaders and followers need to resist interference from jammers during the formation tracking process with a leader-follower structure. Leaders and followers achieve formation tracking through cooperation. The interactions between jammers and the other UAVs form a non-cooperative game. This is then used as the basis of Tri-Level Actor-Critic (Tri-AC) reinforcement learning algorithm for decision-making.

As can be seen, previous research on UAV anti-jamming measures has already explored some methods based on multi-UAV formation control. Our approach differs by proposing a GA algorithm to jointly optimize UAV formation, antenna beam directions, and traffic routing in a way that minimizes the communication disruption caused by the jammer.

\section{System Model}

The proposed model exploits beam-steering antennas to establish
links with less susceptibility to interference and jamming, together with formation positions that favor the establishment of these
directional links. The communication capacity between UAVs is thus influenced by inter-node distance, transmit power, antenna beam direction, and interference from jammers.

At this stage, the proposed model is based on the following assumptions about the communication channels:
\begin{enumerate}
	\item The position of each UAV is defined in 2D Cartesian coordinates (x,y), and its smart antenna beam direction is represented as an angle between 0° and 360°;
	\item The UAV swarm is fully autonomous, with no need for communication with a Ground Control Station (GCS);
	\item The smart antennas allow the estimate of the Angle of Arrival (AoA) of the jamming signal, thus allowing its localization to be estimated by the swarm;
	\item The UAVs seek to maintain the main communication channel, which has a higher data rate, thus being more vulnerable to jamming;
	\item Main channel beam-steering and formation coordination messages are idealistically conveyed through a dedicated omnidirectional low data rate robust signaling channel, which is resilient to jamming;
	\item The optimization algorithm is run in one of the UAVs, which receives the input data from the other UAVs (through the signalling channel), computes the solution, and transmits the solution back to the other UAVs through the signalling channel.
\end{enumerate}

A system model with single communication channel, where the signaling messages are also affected by jamming is relegated for future work. The main objective of this paper is to show how swarm member behavior can mitigate jamming when properly coordinated.

\begin{figure}[!h]
	\centering
	\includegraphics[scale = 0.6]{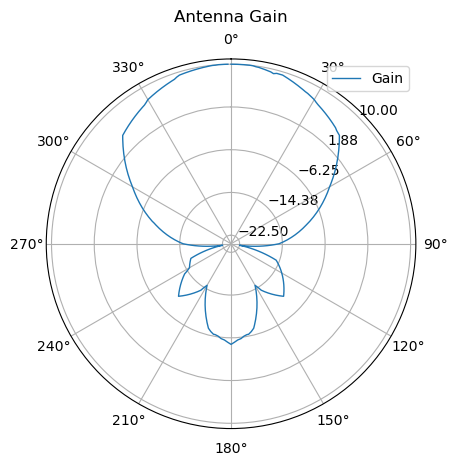}
	\caption{Radiation diagram for the smart antenna.}
	\label{fig1}
\end{figure}

Smart antenna capabilities are based on a real antenna radiation pattern. Based on the radiation diagram, it is possible to accurately calculate transmission and reception gains between each UAV and the jammer. Based on \cite{lee2019tilted}, which examined high-efficiency, 360° directional antennas at 5.9 GHz for UAV radar applications, this study adapted typical gain values from these antennas to create an effective radiation model, which is depicted in Figure \ref{fig1}. 

\begin{figure}[!h]
	\centering
	\includegraphics[scale = 0.55]{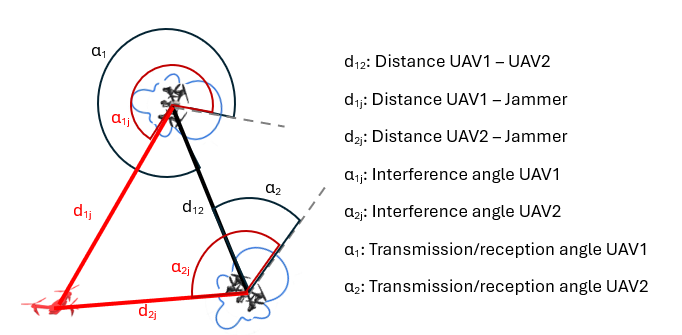}
	\caption{Acquisition of reception/transmission distances and relative angles.}
	\label{DQLscheme}
\end{figure}

Each UAV can freely position itself in 2D space within a designated swarm deployment area, which is discretized into a grid of points (e.g., 5~m spacing between horizontal and vertical neighbor points). Optimization focuses on maximizing link capacities under jammer interference, with antenna directions adjustable from 0 to 360 degrees. UAVs continuously measure and share their positions and movement vectors by means of the signaling channel, enabling calculation of transmission, reception, and interference powers based on transmission/reception angles and distances, as illustrated in Figure \ref{DQLscheme}. 

A log-distance path loss model calculates link capacity in UAV networks, which accounts for various factors impacting communication quality. The path loss $PL$ is determined using Eq. (\ref{Eq12}), where parameters such as loss at a reference distance $PL_0$, the loss exponent $\gamma$, and distance $d$ play crucial roles.

\begin{equation}\label{Eq12}
	PL = PL_0 + 10\gamma \log_{10}\left(\frac{d}{d_0}\right) + X_g
\end{equation}

This loss value is subsequently used to calculate the received power $P_{\text{RX\_dBm}}$ with Eq. \ref{Eq13}, incorporating transmission and antenna gains. The signal-to-noise ratio (SNR) is derived in linear units from the received power and noise power, enabling the calculation of link capacity $C$ using Shannon-Heartley theorem, in Eq. (\ref{Eq15}).
\begin{equation}\label{Eq13}
	P_{r} = P_{t} + \text{G}_{t} + \text{G}_{r} - PL
\end{equation}

\begin{equation}\label{Eq15}
	C = B \cdot \log(1 + \text{SNR})
\end{equation}

Each state of the environment corresponds to a non-symmetrical capacity matrix storing UAV link capacities. The Dijkstra algorithm is applied to identify paths that maximize transmission capacity, treating the system as a graph where edge weights are inversely proportional to link capacities given by $\omega_l = \frac{1}{C_{l}}$,
where $\omega_l$ and $C_l$ are the weight and capacity of link $l$, respectively. This method identifies the lowest-cost paths between UAVs, with the bottleneck value representing the lowest capacity encountered along these paths. Given an end-to-end path between UAVs $i$ and $j$, $p_{i,j}$, encompassing all intermediate links, the capacity in the end-to-end path takes bottlenecks into account as follows:

\begin{equation}\label{E2E_C}
	C^{E2E}_{i,j} = \text{min}({C_{l}}): l\in p_{i,j}.
\end{equation}

 The objective function ($OF$) for this optimization problem is defined in Eq. (\ref{Eq17}), maximizing the weighted product of average capacity $C^{E2E}_{avg}$ and minimum capacity $C^{E2E}_{min}$ of all the $C^{E2E}_{i,j}$, with weights $\alpha$ and $\beta$ allowing for the relative prioritization of each term. This non-linear approximation ensures that variations in capacity metrics do not unduly influence the optimization direction, enabling balanced decision-making based on specific application needs.

\begin{equation}\label{Eq17}
	\text{Maximize} \quad OF(p,\theta)={C^{E2E}_{avg}(p,\theta)}^{\alpha }\cdot{C^{E2E}_{min}(p,\theta)}^{\beta }
\end{equation}

\begin{table}[]
	\centering
	\setlength{\tabcolsep}{3pt} 
	\caption{Adaptable variables for UAV communication system}\label{Tab3}
	\begin{tabular}{@{}>{\raggedright}p{2.0cm} >{\raggedright}p{5.0cm} >{\raggedright\arraybackslash}p{1cm}@{}}
		\toprule
		\textbf{Symbol} & \textbf{Description} & \textbf{Unit} \\ \midrule
		$B$ & Bandwidth of the main communication channel & Hz \\ 
		$P_{j}$ & Transmit power of the jammer signal & dBm \\ 
		$P_{t}$ & Transmit power of each UAV & dBm \\ 
		$N_{UAV}$ & Number of UAVs in the system &  \\
		$L$ & Length of the area of interest & meters \\ 
		$R$ & Radiation pattern of UAV antennas &  \\ 
		$\gamma$ & Path loss exponent  &  \\
		$d_0$ & Reference distance & meters \\
		$PL_0$ & Path loss value at reference distance $d_0$ & dB \\ 
		$T$ & Time limit for the output of the simulation result & seconds \\ \bottomrule
	\end{tabular}
	\label{tab:design-variables}
\end{table}

\section{Genetic Algorithm Approach}

\begin{figure}[!h]
	\centering
	\includegraphics[scale = 0.5]{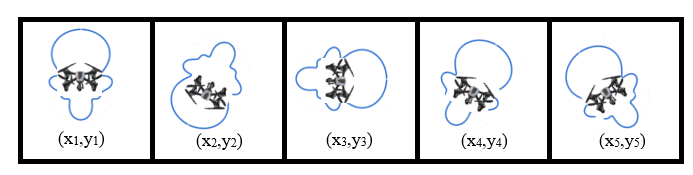}
	\caption{GA chromosome structure. }
	\label{FIG7}
\end{figure}

The chromosome defined for the GA is a sequence of values (genes) that encode the antenna beam direction and 2D position for each UAV. The chromosome is made up of a number of genes equal to the number of UAVs under study. The used chromosome structure is depicted in Figure \ref{FIG7}.

The GA starts by initializing a population of random chromosomes. The fitness values of each chromosome is computed based on the objective function in Eq. (\ref{Eq17}). Through a selection process involving tournament selection (where the best chromosome from each group survives into the next generation), the developed GA encourages favorable formations. 



An encapsulated GA architecture was used. The outer-GA performs Position Optimization, while the inner-GA performs Beam Direction Optimization. Position optimization seeks to attain the optimal spatial arrangement of UAVs. However, the objective function also depends on beam directions. As such, for each set of UAV positions, the innerGA is used to adjust the antenna beam directions to maximize the objective function. This encapsulated GA approach reduces the complexity of the search by focusing on foundational position optimization before fine-tuning radiation pattern directions, thereby enhancing convergence efficiency and improving the final solution quality.

\begin{figure}[!htbp]
	\centering
	\includegraphics[width=0.8\linewidth]{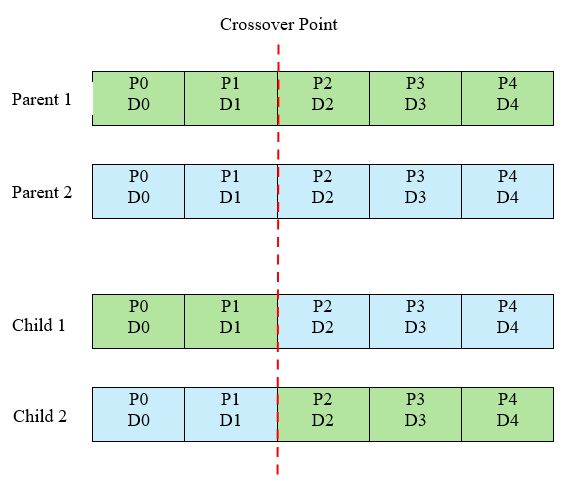}
	\caption{Crossover operator. }
	\label{FIG8}
\end{figure}

Regarding genetic operators, there are two crossover operators, one for outer-GA and another for inner-GA. Both consist of randomly selecting a crossover point in the parent chromosomes and generating the two children by combining one fraction from each parent (UAV positions for outer-GA, antenna beam directions for inner-GA). Figure \ref{FIG8} demonstrates the crossover process.

 There are also two mutation operators are defined, for outer-GA and inner-GA. The outer-GA mutation consists of assigning a new random position to each selected UAV, selecting among the eight closest positions of the grid. The inner-GA mutation undergoes a random deviation of the beam direction of each selected UAV in the range [-20, 20] degrees. \ref{FIG8} demonstrates the mutation process.

\begin{figure}[!htbp]
	\centering
	\includegraphics[width = 0.6\linewidth]{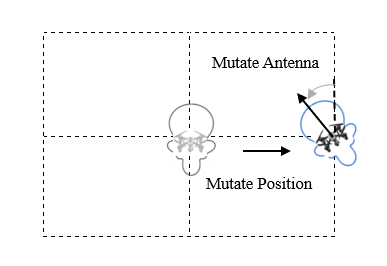}
	\caption{Mutation operator. }
	\label{FIG9}
\end{figure}

Figure \ref{FIG9} shows the mutation process.

\section{Simulation Results}
The simulation software was implemented in Python. Seaborn, a Python data visualization library, was used, since it provides a high-level interface for exploring complex datasets. This library enables effective visualization, interpretation, and comparison of the results obtained in the research.

The main simulation parameters are detailed in Table \ref{Tab6}.

\begin{table}[!h]
	\centering
	\caption{Parameters applied to the UAV swarm}\label{Tab6}
	\begin{tabular}{@{}ll@{}}
		\toprule
		\textbf{Parameter}            & \textbf{Value}      \\ \midrule
		$N_{UAV}$         & 4               
		\\
		B & $2.4e^{9}$ Hz 
		\\
		$P_t$ & 20 dBm
		\\
		$d_{0}$ & 1~m
		\\
		$PL_0$ & 30~dB 
		\\
		
		$P_j$ & 100~dBm
		\\
		$\gamma$ & 2 
		\\
		
		Size of the area of interest & 50~m x 40~m \\
		Spacing between discretized points & 5~m  
		\\ \bottomrule
	\end{tabular}
	\label{Parametros do Enxame}
\end{table}

The results obtained through simulations provide insightful comparisons across the three study scenarios, each with progressively increasing complexity, which showcase the strengths and limitations of the proposed GA approach. The tests were carried out in a laptop computer equipped with an AMD Ryzen 7 6800HS with Radeon Graphics, 3.20~GHz, 16~GB of memory.

\subsection{Scenario 1: Static Positions with Beam-Steering Optimization}
\begin{figure}[!h]
	\centering
	\includegraphics[scale = 0.6]{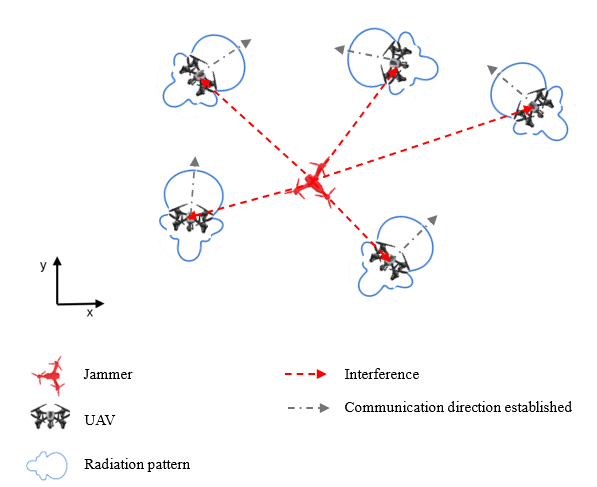}
	\caption{Scenario 1: optimizing antenna beam directions for fixed UAV positions.}
	\label{cen1}
\end{figure}

In a static UAV swarm scenario, the focus is on evaluating the beam-steering of the UAV antennas, so that the interference of the jamming signal is mitigated. Key evaluation factors include the effectiveness of beam alignment for required transmissions, as well as reception efficiency in a static environment. Figure \ref{cen1} illustrates the scenario.

Initially, the parameters that favored the solution found in terms of execution time were studied. The following GA parameters were analyzed: population size, number of generations, mutation rate and crossover rate. The results are listed in Table  \ref{tab:parameters-results}.

\begin{table}[!h]
	\centering
	\caption{Study of GA parameterisation.}
	\begin{tabular}{@{}>{\raggedright}p{1cm} >{\raggedright}p{1cm} >{\raggedright}p{1cm} >{\raggedright}p{1cm} >{\raggedright}p{1cm} >{\raggedright\arraybackslash}p{1cm}@{}}
		
		\toprule
		\textbf{Population Size} & \textbf{Number of Generations } & \textbf{Mutation Rate} & \textbf{Crossover Rate} & \textbf{Objective Function Value} & \textbf{Simulation Time[s]} \\ \midrule
		10 & 20 & 0.15 & 0.8 & 15810 & 7 \\
		10 & 20 & 0.15 & 0.9 & 31218 & 8 \\
		10 & 50 & 0.15 & 0.8 & 38490 & 10 \\
		10 & 100 & 0.15 & 0.8 & 72333 & 16 \\ \midrule
		50 & 10 & 0.15 & 0.9 & 135466 & 10 \\
		50 & 50 & 0.15 & 0.9 & 197841 & 13 \\
		50 & 100 & 0.15 & 0.9 & 329988 & 41 \\
		50 & 200 & 0.15 & 0.9 & 364068 & 73 \\ \midrule
		100 & 10 & 0.15 & 0.9 & 204256 & 14 \\
		100 & 50 & 0.15 & 0.9 & 394567 & 27 \\
		100 & 100 & 0.15 & 0.9 & 408590 & 64 \\
		100 & 200 & 0.15 & 0.9 & 402333 & 143 \\ \bottomrule
	\end{tabular}
	\label{tab:parameters-results}
\end{table}

The value of the objective function has converged in all tests. However, increasing the number of generations will delay the response. In a realistic setting, response time will be relevant to the success of the swarm. Table \ref{tab8} shows the default parameters that were selected for the GA in the remaining simulations.

\begin{table}[!h]
	\centering
	\caption{Parameters used in GA}
	\begin{tabular}{@{}ll@{}}
		\toprule
		\textbf{Parameter Value}            & \textbf{Value}      \\ \midrule
		Population Size         & 100               
		\\
		Generations                & 50                 \\
		Mutation rate              & 15\%                 \\
		Crossover rate             & 90\%                 \\ \bottomrule
	\end{tabular}
	\label{tab8}
\end{table}

\begin{figure}[!h]
	\centering
	\includegraphics[width = \linewidth]{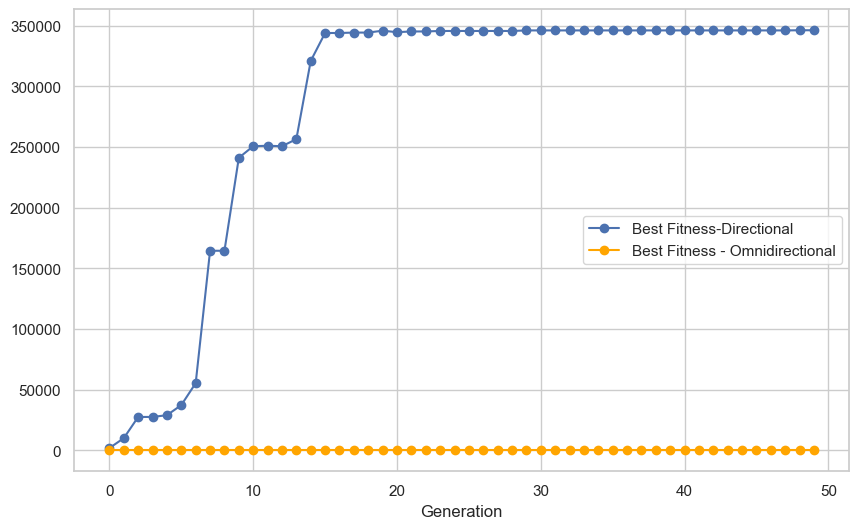}
	\caption{Scenario 1: convergence of objective function values with GA. }
	\label{Fig16}
\end{figure}

Figure \ref{Fig16} shows the evolution of the OF value for a beam-steering configuration, as well as the constant baseline value corresponding to the use of omnidirectional communication. The figure shows that the application of beam-steering antennas can effectively improve the performance of communications: the best fitness value for the omnidirectional configuration is 0.11, in contrast with the value obtained for the
beam-steering configuration, which is approximately 350000.

\begin{figure}[!h]
	\centering
	\includegraphics[width = \linewidth]{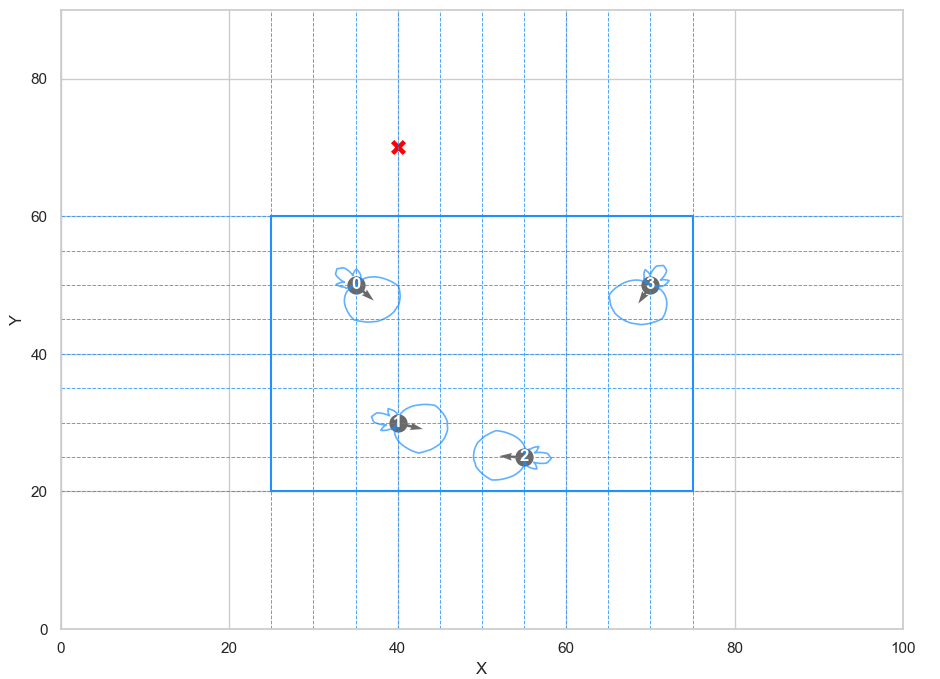}
	\caption{Scenario 1: Beam-steering optimized by the GA. }
	\label{fig17}
\end{figure}

A specific example is depicted in Figure \ref{fig17}, which was obtained after a computation time of 29.3~s. This result corresponds to an average data rate of 2856~bps and a bottleneck data rate of 129~bps.  In contrast, for the omnidirectional configuration, the average data rate was 0.89~bps and the bottleneck data rate was 0.12~bps. Figure \ref{fig17} shows the direction of the antenna taken by each UAV. It is easy to notice that the beam-steering solution favors pointing the beams away from the jammer (e.g., UAV0 and UAV3), and/or towards other UAVs (e.g., UAV1 and UAV2).

\subsection{Scenario 2: Fixed Swarm Deployment Area with Beam-Steering and UAV Position Optimization}
\begin{figure}[!h]
	\centering
	\includegraphics[width = \linewidth]{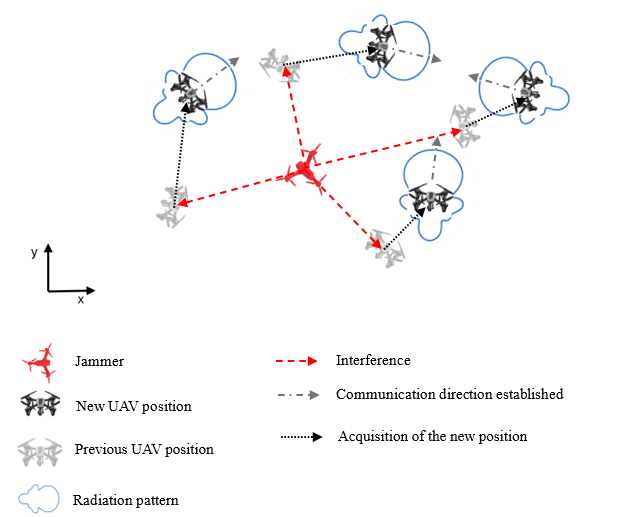}
	\caption{Scenario 2: exploitation of antenna directionality and UAV positions.}
	\label{cen2}
\end{figure}

In this scenario, both UAV swarm spatial positioning and beam-steering are considered, allowing individual UAVs to adjust positions relative to each other, within a statically defined deployment area. Figure \ref{cen2} illustrates the swarm capabilities in this scenario.

\begin{figure}[!h]
	\centering
	\includegraphics[width = \linewidth]{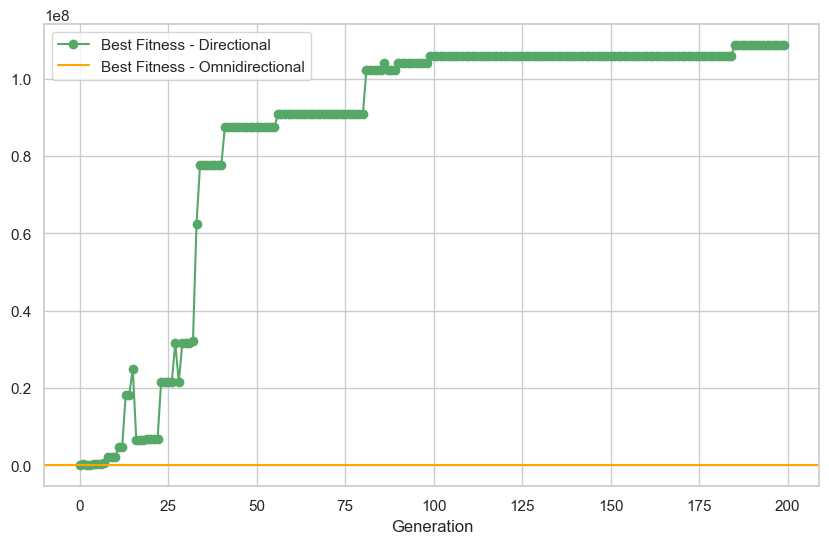}
	\caption{Scenario 2: convergence of the OF in acquiring the positions.}
	\label{Fig10}
\end{figure}

Figure \ref{Fig10} shows that, given the lower complexity of the possible positions, the GA converges more stably and quickly. There is again a significant difference between the attained OF value when applying beam-steering (1.08$e^{8}$), compared with the omnidirectional case (1587). 

\begin{figure}[!htbp] 
	\centering
	\subfloat[Solution obtained after 15~s of computation.\label{fig:sub1}]{%
		\includegraphics[width=\linewidth]{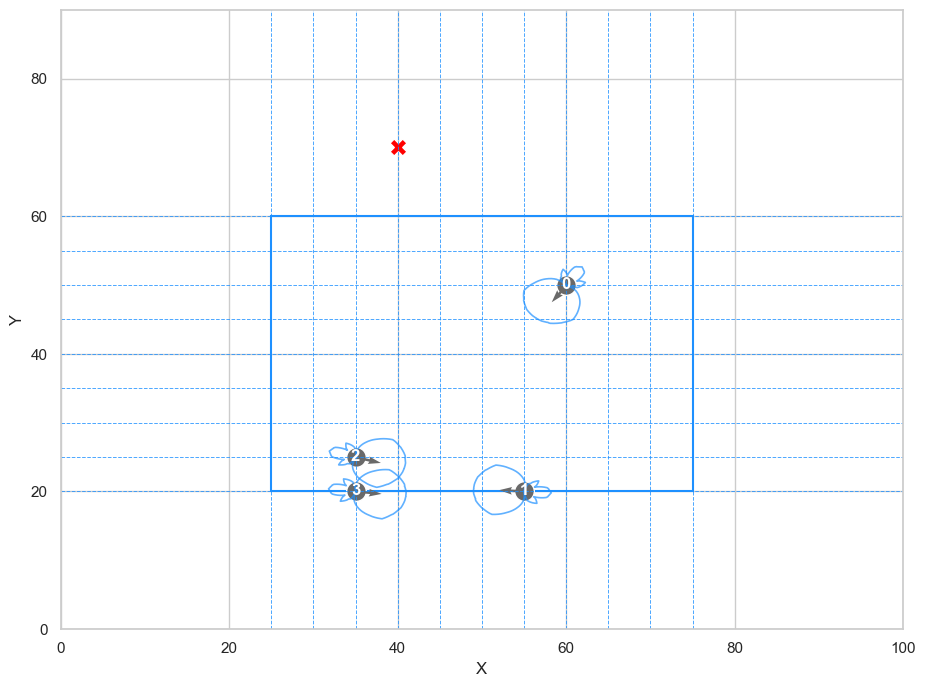}%
	}
	\par 
	\subfloat[Solution obtained after 60 seconds of computation.\label{fig:sub2}]{%
		\includegraphics[width=\linewidth]{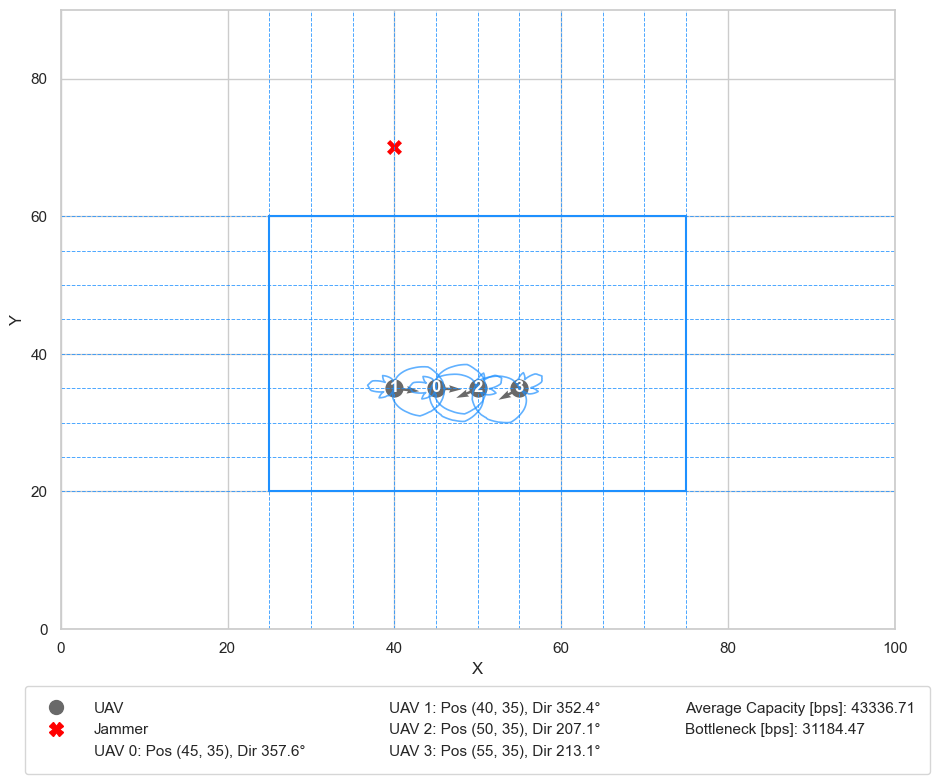}%
	}
	\caption{Scenario 2: GA solutions obtained for two different computation times for the beam-steering configuration.}
	\label{fig28}
\end{figure}

The scenario was also analyzed for two different computation time limits: 15~s and 60~s. The obtained results are shown in Figure \ref{fig28}.

\begin{figure}[!htbp]
	\centering
	\includegraphics[width=\linewidth]{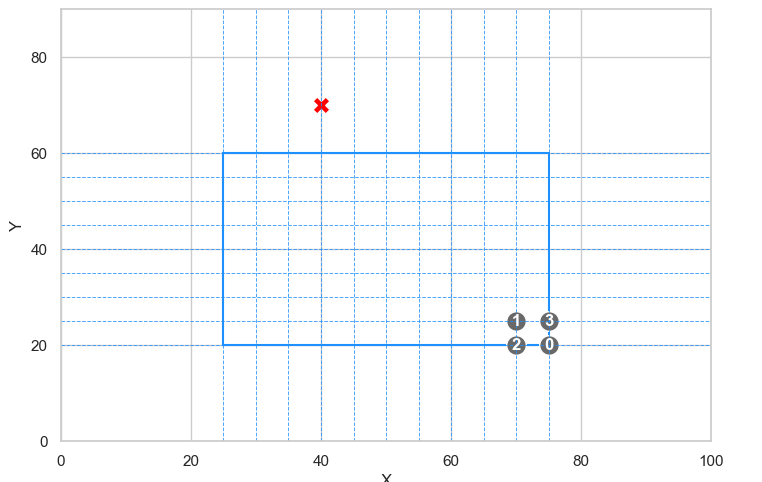}
	\caption{Scenario 2: GA results obtained after 23 seconds of computation with omnidirectional antennas.}
	\label{fig29}
\end{figure}

In the omnidirectional communication configuration, GA can obtain a very fine-tuned arrangement of the UAVs in a significantly shorter time interval, trying to place the swarm as far as possible from the jammer. Its results are shown in Figure \ref{fig29}.
 Although the positioning of the UAVs using the beam-steering antenna configuration are obviously not ideal when compared to the omnidirectional configuration, there is a drastic decrease in communication performance when using omnidirectional antennas.

Analyzing the results above, it is clear that the GA seeks to acquire positions further away from the jammer to reduce the interference felt, and, on the other hand, it aims to bring the UAVs closer together to increase the power received by each node. The results improve for longer could be better since, to increase the objective function value, the GA requires a computation time interval that may not compatible with a realistic operating environment. This issue is addressed in Scenario 3.

\subsection{Scenario 3: Mobile Swarm Deployment Area}
\begin{figure}[!htbp]
	\centering
	\includegraphics[width = \linewidth]{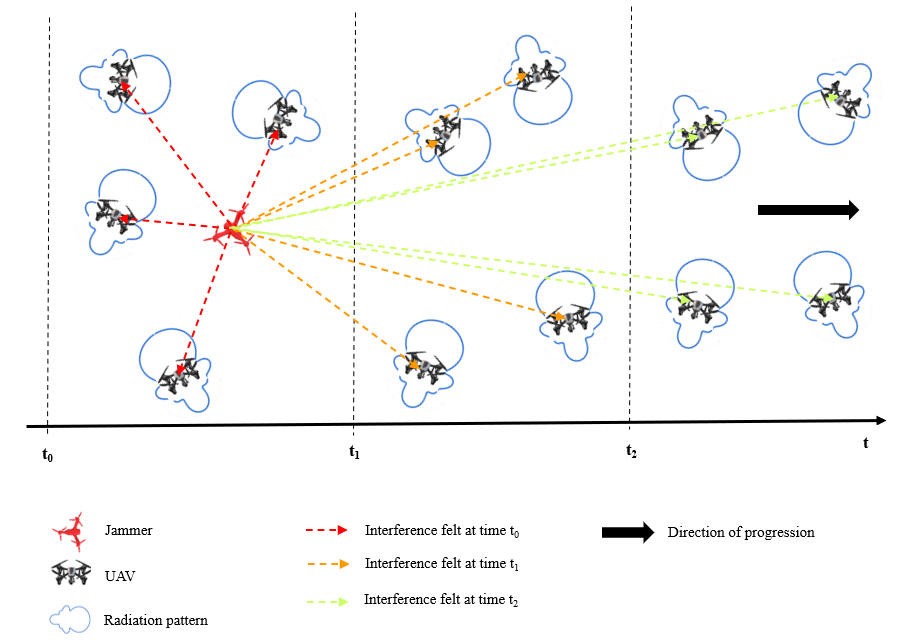}
	\caption{Scenario 3: UAV swarm moves horizontally.}
	\label{cen3}
\end{figure}

\begin{figure}[!htbp]
	\centering
	\includegraphics[width = \linewidth]{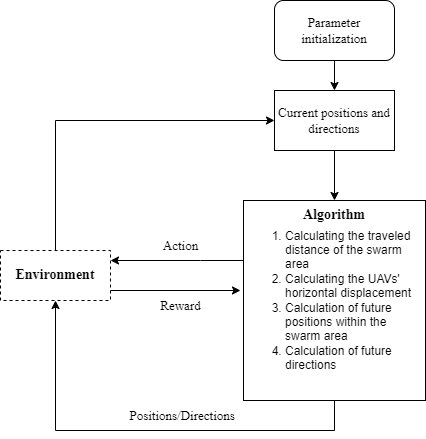}
	\caption{Scenario 3: simulation execution flowchart. }
	\label{figura_1}
\end{figure}

In this dynamic scenario, the UAV swarm continuously moves in a straight line towards some mission objective, while the UAVs can change their relative positions within its deployment area, as well adapt the beam-steering configuration accordingly. Figure \ref{cen3} illustrates the scenario. Figure \ref{figura_1} shows the system execution flowchart. Additional simulation parameters for Scenario~3 are listed in Table \ref{Swarm parameters}.

\begin{table}[!htbp]
	\centering
	\caption{Scenario 3: additional parameters of the UAV swarm.}
	\begin{tabular}{@{}ll@{}}
		\toprule
		\textbf{Parameter} & \textbf{Value} \\ \midrule
		GA refresh interval & 7.5 s 
		\\
		UAV speed & 2 m/s 
		\\
		Direction of progression & Axis of xx
		
		\\ \bottomrule
	\end{tabular}
	\label{Swarm parameters}
\end{table}

Simulation was performed assuming discretization of time in windows of 7.5~s. During each time window, the UAV swarm has to recalculate the formation and beam-steering configuration for the next time window, which corresponds to the optimization task in Scenario~2. In order to meet this constraint, the GA was parameterized with a small population of 14 chromosomes and maximum of 40 generations, which produces results within an average simulation time of 6.43 seconds. 

\begin{figure}[!htbp]
	\centering
	\includegraphics[width=0.8\linewidth]{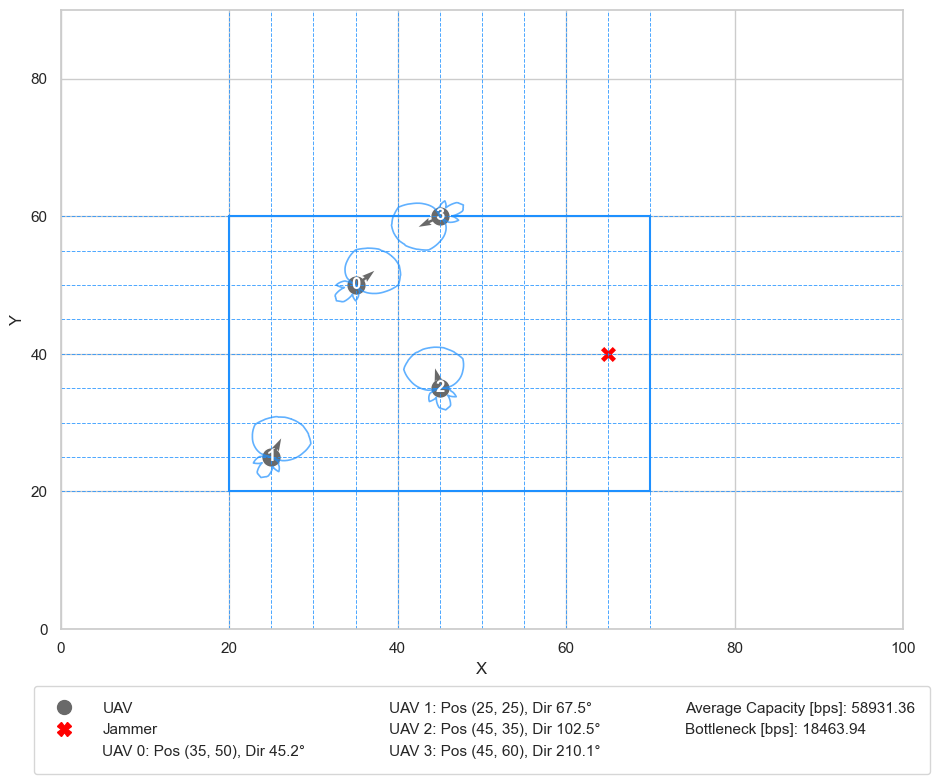}
	\caption{Topology at simulation time $0.0s$. }
	\label{Fig35}
\end{figure}

The initial UAV swarm situation is depicted in Figure \ref{Fig35}. 

\begin{figure}[!htbp]
	\centering
	\begin{subfigure}[b]{0.40\textwidth}
		\centering
		\includegraphics[width = \linewidth]{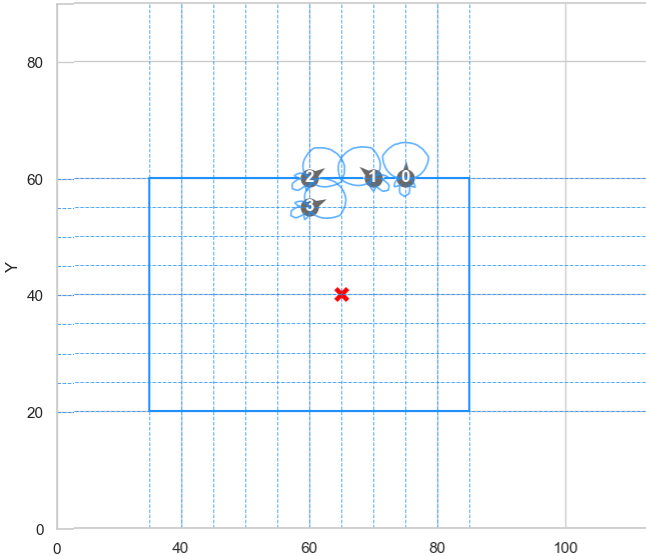}
		\caption{Topology at simulation time $7.5s$.}
		\label{fig:image2}
	\end{subfigure}
	\vfill
	\begin{subfigure}[b]{0.40\textwidth}
		\centering
		\includegraphics[width = \linewidth]{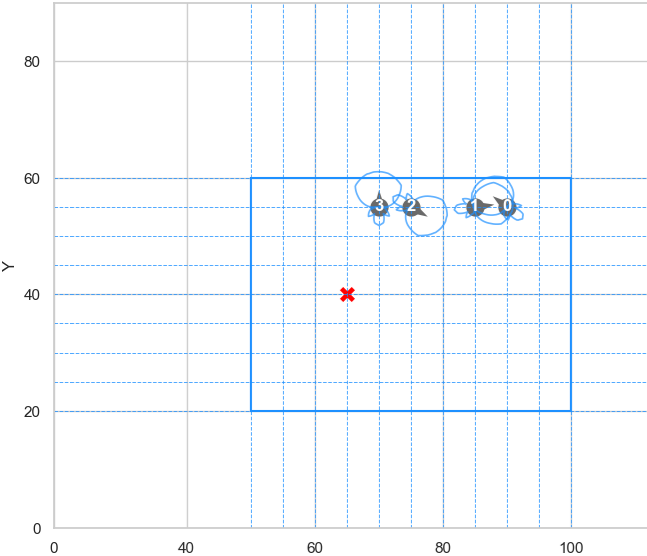}
		\caption{Topology at simulation time $15.0s$.}
		\label{fig:image3}
	\end{subfigure}
	\caption{Scenario 3: GA formation and beam-steering adaptation as the UAV swarm moves horizontally.}
	\label{Fig37}
\end{figure}

Selected snapshots of the UAV swarm movement are depicted in Figure \ref{Fig37}. Despite processing time constraints, the GA could effectively cluster UAVs and direct antennas to mitigate jamming effects. Over the simulation, the sequence of objective function curves (omitted due to page limits) reflect a steady optimization of communication quality, as seen in evolving UAV topologies as the swarm operational area slides horizontally. The success in aligning antennas and adjusting positions demonstrates the GA capability to provide robust anti-jamming strategies within the given constraints, even if the found solutions were still not optimal. It must be noted that the GA seeks optimal UAV positions and beam-steering solutions for specific time instants of the swarm movement. Although the GA could improve its decisions for longer time windows, the latter would also mean that the swarm would adapt its topology and beam-steering less frequently. For very distant jammers, enlarging the time window is possible, since angular directions change slowly, otherwise, enlarging the time window will also lead to non-optimality. Optimization of the time window length is left for future work.

\section{Conclusions}
This paper has presented a GA based method to optimize beam-steering and UAV positions to attain improved anti-jamming. Three scenarios were used to evaluate the proposed solution. 

In the first scenario, the UAV positions remain static, and only the directions of the directional antennas are adjusted. The results show the effectiveness of GA to perform beam-steering in an optimal way. It also allowed to confirm the significant anti-jamming performance improvement when using beam-steering, relative to omnidirectional antenna operation.

In second scenario, the degree of complexity was significantly increased with the introduction of an additional degree of freedom for the individual movement of each UAV. Although it is considered that the deployment area of the swarm is fixed, UAVs are allowed to adapt their positions within this deployment area. The results show that the effectiveness of the GA increases with the computation time available for acquiring the optimal formation, which constitutes a constraint in more dynamic and realistic scenarios.

Finally, the third scenario considers a moving UAV swarm. Time discretization in a sequence of time windows, makes it equivalent to a sequence of instances of the second scenario. In this scenario, available GA computation time is constrained by the duration of the time window. The results show that, although the GA is able to find acceptable solutions in each time window, the time window constraint leads to non-optimality. 

Ongoing work by this team is currently focused on reducing response time by means of Machine Learning techniques, in particular Deep Reinforcement Learning. Other aspects that are currently under test are the use of antennas with different radiation patterns, as well as improved models of anti-jamming performance taking into account intermediate steps of the movement. The proposed GA solution is inherently centralized. Future work will focus on the use of decentralized approaches, such as Multi-Agent Reinforcement Learning (MARL). By having each UAV making its own decisions learned in a collaborative multi-agent environment, this approach may allow to drop the assumption of a jamming-free signaling channel, which can fairly be considered unrealistic. 

\bibliographystyle{IEEEtran}
\bibliography{referencias}  

\end{document}